\documentclass[useAMS,usenatbib,usegraphicx]{mn2e}

\newcommand{\cii}{{\sc [C\,ii]}}
\newcommand{\lcii}{\ensuremath{L_{\mathrm{[C\,II]}}}}
\newcommand{\lfir}{\ensuremath{L_{\mathrm{FIR}}}}
\newcommand{\ratio}{\ensuremath{\lcii / \lfir}}

\title[An Upper Limit to {[C\,{\normalsize \it II}]} Emission in a $z
\simeq 5$ Galaxy]{An Upper Limit to [C\,{\Large\bf II}] Emission in a
$\bmath{z \simeq 5}$ Galaxy}

\author[G. Marsden et al.]{
\parbox{159mm}{
\begin{flushleft}
Gaelen Marsden$^{1}$\thanks{E-mail:gmarsden@physics.ubc.ca},
Colin Borys$^{2}$,
Scott C. Chapman$^{2}$,
Mark Halpern$^{1}$ and
Douglas Scott$^{1}$
\end{flushleft}
}\vspace*{0.200cm}\\
\parbox{159mm}{
$^{1}$Dept. of Physics \&
Astronomy, University of British Columbia, 6224 Agricultural Rd.,
Vancouver BC, V6T 1Z1, Canada\\
$^{2}$California Institute of Technology, 1201 East California Blvd.,
Pasadena CA, 91125, USA}}

\begin{document}

\pagerange{\pageref{firstpage}--\pageref{lastpage}} \pubyear{2003}

\maketitle

\label{firstpage}

\begin{abstract}
Low-ionization-state far-infrared (FIR) emission lines may be useful
diagnostics of star-formation activity in young galaxies, and at high
redshift may be detectable from the ground. In practice, however, very
little is known concerning how strong such line emission might be in
the early Universe. We attempted to detect the $158\,\micron$ \cii\
line from a lensed galaxy at $z=4.926$ using the Caltech Submillimeter
Observatory. This source is an ordinary galaxy, in the sense that it
shows high but not extreme star formation, but lensing makes it
visible. Our analysis includes a careful consideration of the
calibrations and weighting of the individual scans. We find only
modest improvement over the simpler reduction methods, however, and
the final spectrum remains dominated by systematic baseline ripple
effects. We obtain a 95 per cent confidence upper limit of $33\,$mJy
for a $200\,$km$\,$s$^{-1}$ full width at half maximum line,
corresponding to an unlensed luminosity of $1 \times 10^9\,$L$_\odot$
for a standard cosmology. Combining this with a marginal detection of
the continuum emission using the James Clerk Maxwell Telescope, we
derive an upper limit of 0.4 per cent for the ratio of \ratio\ in this
object.
\end{abstract}

\begin{keywords}
galaxies: high redshift -- galaxies: evolution -- galaxies:
formation -- cosmology: observations -- submillimetre
\end{keywords}

\section{Introduction}

Low-ionization-state far-infrared (FIR) emission lines play an
important role in cooling star-forming regions, and they allow us to
infer the flux of rest-frame ultraviolet (UV) photons. Star-forming
galaxies have large dust masses which obscure UV radiation, so we rely
on the IR and longer wavelengths to probe these star-forming
regions. Studies have shown that \cii, the ground-state fine-structure
line of singly ionized carbon (${^2P}_{3/2} \rightarrow {^2P}_{1/2}$,
$\lambda = 157.7409\,\micron$), is the dominant cooling line in
gas-rich star-forming regions \citep{dal72,cra85,sta91}. It is thus
expected to be a probe of star formation in young galaxies. In some
nearby galaxies the \cii\ line accounts for as much as 1\% of the
far-IR luminosity \citep{sta91,nik98,moc00}. On the other hand, in
some more distant ultraluminous infrared galaxies (ULIRGs) results
from {\em ISO} indicated much lower fractions
\citep{mal01,con02,luh03}. At $z \simeq 5$ we know very little
concerning the physical conditions within star-forming
galaxies. Metallicity, UV flux, dust content and geometry could all be
radically different from that for more nearby (and hence cosmically
older) systems. In the absence of much empirical data, even upper
limits on \cii\ luminosities are therefore worthwhile.

\cii\ is the brightest line in our Galaxy \citep{miz94}, but its
wavelength makes it impossible to observe from the ground. At high
redshift, where the line is redshifted into the submillimetre, there
is great promise for using this line to probe early galaxy evolution,
or even to find the first objects (e.g.\ \citealt{sta97,sug98}). Until
instruments with much wider back-ends and greatly improved sensitivity
become available, progress can only be made slowly, by targetting
individually promising objects.

The $z = 4.926$ lensed galaxy, detected as a red arc in the cluster
CL~1358+62 \citep[referred to as `G1' in][hereafter F97]{fra97},
magnified by a factor of $5-11$, represents a particularly promising
target for observing \cii\ from the ground. Its redshift places the
line at an accessible frequency ($\simeq 320$\,GHz), though not ideal
since it is on the edge of an atmospheric water line. The galaxy has a
reasonably high inferred star formation rate ($\simeq
36\,$M$_\odot\,$yr$^{-1}$, F97), and unlike quasars observed at
similar redshifts, the \cii\ line is expected to be close to the
systemic redshift of the galaxy. Moreover, although rest-frame UV
colours indicate significant reddening, there is little submillimetre
continuum detected, with a $3 \sigma$ upper limit of $4\,$mJy
(\citealt{vdw01}, but see Section~\ref{res}).

In the following paper we discuss the analysis and interpretation of
our data. Section~\ref{obs} briefly describes the observing
programme. Section~\ref{anlsis} describes the specifics of our analysis
procedures. The results are presented in Section~\ref{res} and
discussed in Section~\ref{dis}.

\section{Observations}
\label{obs}

Object G1 was targetted with the Caltech Submillimeter Observatory
(CSO), located at 13\,300 feet on the summit of Mauna Kea in Hawaii,
over three nights in 1998 January. The telescope is a 10.4\,m {\em
f}/0.4 submillimetre antenna with an alt-az mount.

The atmospheric transmission during the observing run, as measured by
the CSO $\tau$ meter at 225\,GHz, was very good. The optical depth
ranged from $\tau_{225} \simeq 0.03$ to 0.07, corresponding to
optical depth at 320\,GHz of 0.32 to 0.55.

\subsection{Frequency Centre}
\label{obs_freq}

F97 determined the redshift of G1 from the detection of Ly\,$\alpha$
emission at 7204\,\AA, giving $z = 4.926$. They note, however, that
the Ly\,$\alpha$ line is assymetric and that the measured Si\,{\sc ii}
absorption line is blue shifted by $\sim 300$\,km\,s$^{-1}$. These
features can be explained by an outflow model, whereby an expanding
shell of neutral and ionized material absorbs the blue side of the
Ly\,$\alpha$ line (e.g.\ \citealt{leq95}). We assume that the \cii\
emission would be somewhere between the Ly\,$\alpha$ and Si\,{\sc ii}
lines, and therefore centred the detector shifted $\Delta v \approx
200\,$km\,s$^{-1}$ from the Ly\,$\alpha$ centre, or $z = 4.922$.  This
is a reasonable assumption, as this shift is typical of linewidths
observed in local galaxies.  The \cii\ emission line is redshifted to
$\nu_0 = 320.939$\,GHz for the expected centre of our source.

\subsection{Instrumentation}

Based on the calculations of the previous section, it is clear that
the CSO 345-GHz receiver is the appropriate detector. This receiver is
a single side-band SIS mixer with 1\,GHz IF, and acousto-optical
spectrometer (AOS) backends. For our observations the signals from the
mixer were sent to three spectrometers simultaneously, with 50-, 500-,
and 1500-MHz bandwidths.

From the F97 detection of Si\,{\sc ii} in CL\,1358+62-G1, we assume
the \cii\ line is $\simeq 200$\,km\,s$^{-1}$ wide, corresponding to
$200$\,MHz at our observing frequency. We therefore concentrate our
efforts on data from the 1500-MHz receiver, use the 500-MHz data
as a check and ignore the 50-MHz data.

On the first night the receivers were centred on the nominal line
centre, 320.939\,GHz. In an attempt to avoid confusing any systematic
detector response with the shape of the emission line, the receiver
was shifted by $\Delta v = 129.8$\,km\,s$^{-1}$ for the second and
third nights.

The secondary mirror chop throw was set to 60\,arcsec at 1.123\,Hz on
the first night. It was apparent that after observations of one night
we had reached near the instrumental noise limit, so we set more
conservative chop parameters, 40\,arcsec at 0.7\,Hz, on the second and
third nights in an attempt to probe systematic effects. By changing
the chop throw, we additionally decrease the probability of
inadvertantly chopping on to a nearby source.

\subsection{Pointing and Calibration}

The telescope was pointed at $\alpha =
13^\mathrm{h}59^\mathrm{m}39\fs0$, $\delta =
62\degr30\arcmin47\arcsec$. This position was determined by visually
comparing the {\sl Hubble Space Telescope} ({\sl HST}) image presented
in F97 with coordinates of cluster members published by \citet{lup91},
as F97 did not publish coordinates for G1. This process is accurate to
within a few arcsec, which is well within the CSO beamsize at this
frequency ($\simeq 20$ arcsec). Pointing calibrations were performed
occasionally throughout each night, using the
$^{12}$CO\,($3\rightarrow2$) line of IRC\,10216.

Temperature and frequency calibrations were also performed several
times throughout each observing session. Hereafter, `calibration scan'
refers to the temperature calibrations, i.e.\ spectra of a hot load,
while `source scan' refers to spectra of our source, G1.

In all, $\sim 450$ usable source scans were acquired, simultaneously in
each of the 500 and 1500\,MHz detectors. Exact numbers are given in
Table~\ref{obstab}. The 1\,$\sigma$ RMS per raw scan is $\sim 40 -
80$\, mK.

\begin{table}
\caption[Details of observations]{Details of observations. Row 1
shows the line centre of the 
receiver on each night. The rest of the table displays the number of
usable scans taken on each night in each detector.\label{obstab}}
\begin{center}
\begin{tabular}{llrrr}
\hline
\multicolumn{2}{l}{\bf Night:} & \multicolumn{1}{c|}{\bf 1} &
\multicolumn{1}{c|}{\bf 2} & \multicolumn{1}{c|}{\bf 3} \\
\multicolumn{2}{|l|}{Receiver centre (GHz):} & 320.939 & 320.800 & 320.800
\\ \hline 
1500\,MHz & Source scans: & 142 & 236 & 42 \\ 
          & Calibr. scans:   &  42 &  71 & 10 \\ \hline
500\,MHz  & Source scans: & 142 & 236 & 42 \\ 
          & Calibr. scans:   &  41 &  75 & 10 \\ \hline 
\end{tabular}
\end{center}
\end{table}

\section{Analysis}
\label{anlsis}

CSO spectroscopic data are usually analysed using the {\sc class}
(Continuum Line Analysis Single-dish Software, \citealt{bui95})
software package. In the expectation that any detection would be at
very low signal-to-noise ratio, we developed our own analysis
procedures in order to maximise our control over the reduction
process. In this section we describe our methods of calibration and
reduction.

\subsection{Calibration}

CSO data are calibrated in real time by referring each source scan to
the previous hot load calibration scan. Visual inspection of the 500-
and 1500-MHz calibration scans show apparently unstable baselines, so
we investigate whether a more careful calibration will result in lower
noise.

\begin{figure}
\includegraphics[width=84mm]{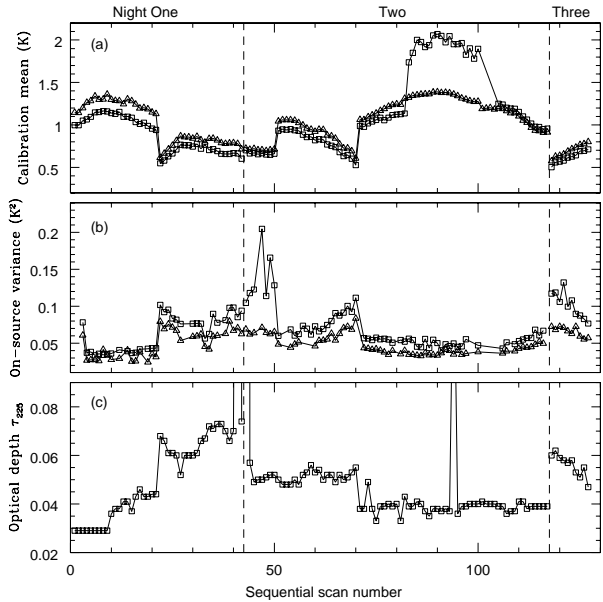}
\caption{Trends in data as a function of time: (a) is the mean value
of each calibration scan for both the 1500- (squares) and 500-MHz
(triangles) spectrometers, measured in K; (b) is the variance of the
source scans for both spectrometers, in K$^2$; (c) is the optical
depth measured at 225\,GHz with the CSO $\tau$ meter. Note the
correlation between all three panels. The vertical dashed lines
indicate the divisions between the three nights.}
\label{calplot}
\end{figure}

We compare the mean value of each calibration scan to the variance of
the following source scans and to the atmospheric opacity,
$\tau_{225}$. We find that the 500- and 1500-MHz calibration means and
the source scan variances all correlate well, and that they track
$\tau$ (see Fig.~\ref{calplot}). 

We propose three possible methods of calibrating the data: (i) for
each source scan, use the most recent calibration scan (the CSO
default); (ii) naively average together all calibration scans within
each night; and (iii) interpolate each calibration channel in time
across each night. The data are thus reduced, as described in
Section~\ref{reduct}, using each of the three calibration
methods. Taking the variance in the final reduced scan as a measure,
we find that method (ii) is clearly inferior to (i), while (i) and
(iii) are not significantly different. We choose method (iii) as the
most elegant solution. If baseline variations did not dominate the
error budget, we might expect method (iii) to yield the best results,
and we would advocate its use in general for low signal-to-noise ratio
data.

\subsection{Data Reduction}
\label{reduct}

The data are co-added within each night. Firstly, a zeroth-order
baseline, fit to the wings of the spectrum, is subtracted from each
source scan in order to minimize the effects of baseline drift. The
data are then combined as a weighted average in each channel. Because
of small shifts in the velocity centres and spacing throughout each
night, we interpolate the data to a common set of velocity bins before
averaging.

The summed scans for each night are binned into roughly
40\,km\,s$^{-1}$ wide channels, and the nights are averaged, weighted
by the variance in each bin. We rebin the data to six different bin
sizes and centres, and find no significant difference in the final
result due to the choice in binning.

\begin{figure}
\includegraphics[width=84mm]{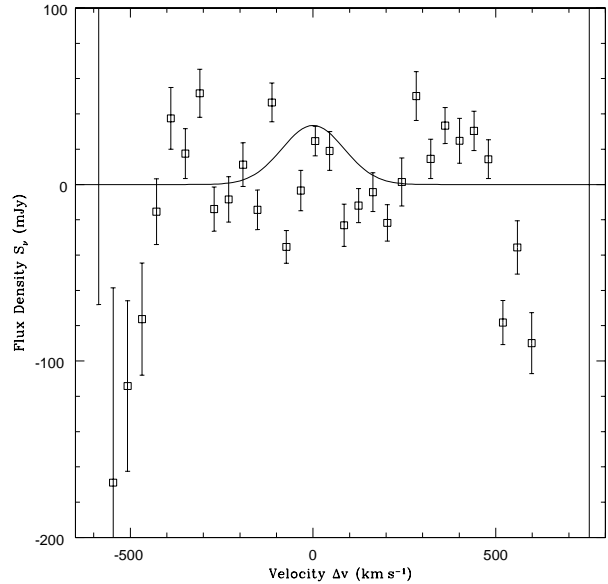}
\caption{The co-added source spectrum, using data from the 1500-MHz
detector. Error bars indicate statistical errors, calculated as the
standard deviation of the data in each bin. The spectrum is dominated
by baseline ripples. The solid curve is the 95 per cent upper limit
for the \cii\ emission line centred at $v = 0$\,km\,s$^{-1}$
(320.939\,GHz) and with FWHM$ = 200$\,km\,s$^{-1}$.}
\label{fluxplot}
\end{figure}

Finally, the co-added spectrum is converted from antenna temperature
to flux density. The beam size and efficiency for the 345-GHz receiver
are 24.6\,arcsec and 74.6 per cent, respectively, \citep{koo01}. The
resulting spectrum is shown in Fig.~\ref{fluxplot}.

\section{Results}
\label{res}

It is clear from Fig.~\ref{fluxplot} that the error bars are much
smaller than the structure in the final spectrum. The systematics in
the instrument prevent integrating down to such low noise levels. We
re-estimate the error bars by finding the best-fit Gaussian and
scaling the errors so that $\chi^2 = N$, where $N$ is the number of
degrees of freedom in the fit. 

A full four-parameter Gaussian fit will not converge to a meaningful
result, so we instead fix the line centre and width to the a priori
expected values of $v_0 = 0$\,km\,s$^{-1}$ and full width at half
maximum (FWHM) = 200\,km\,s$^{-1}$ (as described in
Section~\ref{obs_freq}), and calculate the baseline as the weighted
average of the data outside $\pm 2\sigma$ of the line centre. We find
a 95 per cent upper limit of 33\,mJy using the 1500-MHz receiver; the
500-MHz receiver gives consistent, although less constraining results.
To convert from specific flux $S_\nu$ to luminosity $L$, we first
integrate over frequency, $S = \int S_\nu\,d \nu
=1.06\,S^\mathrm{peak}_\nu \times \mathrm{FWHM}$ for a Gaussian of
height $S^\mathrm{peak}_\nu$ and width $\mathrm{FWHM} = \nu_0 (\Delta
v / c)$, then multiply by $4 \pi D_\mathrm{L}^2$, where $D_\mathrm{L}$
is the luminosity distance for a given cosmology. For a standard
cosmological model with $\Omega_\mathrm{M} = 0.3$, $\Omega_\Lambda =
0.7$, and $H_0 = 70$\,km\,s$^{-1}$\,Mpc$^{-1}$, we find $D_\mathrm{L}
= 45.8\,$Gpc and $\lcii = 5.4 \times 10^9$\,L$_\odot$.

We also rereduce exisiting SCUBA observations at 850\,$\micron$
\citep{hol99} to find an estimate of the total FIR luminosity,
\lfir. An 850-$\micron$ upper limit of $S_\nu < 4$\,mJy was reported
by \citet{vdw01} based on 4\,ks of pointed photometry data. We combine
15\,ks of jiggle-map data ($\sim 2.5$\,mJy rms) with the photometry
data ($\sim 1.5$\,mJy rms) using the methods described in
\citet{bor03} and find a $2.1 \sigma$ peak 2\,arcsec from G1, well
within the 15-arcsec beam.  Although not highly significant, and in a
blank field survey we would not consider a $2\sigma$ peak a detection,
this is a pointed measurement of a known source, so we consider this a
measurement rather than simply an upper limit. We can then calculate
the ratio \ratio\ using Gaussian probability distributions to describe
both \lcii\ and \lfir.  Integrating a grey-body model with
$T_\mathrm{dust} = 40$\,K and $\beta = 1.5$ \citep{com99,kla01} from 8
to 1000\,$\micron$, we then find that the SCUBA measurement $S_{850} =
(2.8 \pm 1.3)$\,mJy corresponds to $\lfir = (2.4 \pm 1.2) \times
10^{12}$\,L$_\odot$, assuming the same cosmology as above. Integrating
over the joint probability, we find a Bayesian 95 (99) per cent upper
limit $\ratio < 0.4$ (1.1) per cent.

\section{Discussion}
\label{dis}

Typical values of \ratio\ are thought to be $\sim 0.1 - 1$ per cent,
but there is recent evidence to suggest that it can be lower,
especially for ULIRGs, which are thought to make up a large fraction
of the high-redshift submillimetre population. We find $\ratio < 0.4$
per cent at 95 per cent confidence for our particular target. While
this does not rule out the ratios found by \citet{sta91}, it does
confine the ratio for this particular high-redshift galaxy to the
lower end of the range.

Based on their lensing model, F97 determine that the source G1 is
magnified by a factor of $5-11$. Taking the lower end of this
magnification range, we find an unlensed luminosity of $\lcii < 1
\times 10^9$\,L$_\odot$.  We can use the absolute \lcii\ measurement
to place an upper limit on the star formation rate (SFR) occurring in
the source. \citet{bos02} relate SFR to \lcii\ by assuming an initial
mass function, with slope $\alpha = 2.35$ between
0.1 and 100\,M$_\odot$, and using their \cii\ --
H$\alpha$ relation determined from observations of a sample of Virgo
galaxies, finding

\[
\mathrm{SFR} = 1.73 \times 10^{-6} \times \left( \lcii / L_\odot
\right)^{0.788} \; \mathrm{M}_\odot\,\mathrm{yr}^{-1} .
\]

Assuming this relation holds for all galaxies at all redshifts, and
that the source is not differentially lensed, we find $\mathrm{SFR}
\la 21$\,M$_\odot$\,yr$^{-1}$. However, we note that \citet{bos02}
claim the relation only holds up to $10^{10.5}\,\mathrm{L}_\odot$ and
that the SFR is uncertain by up to a factor of 10, so we should not
place too much confidence in the estimate. As a comparison, F97 find
$\mathrm{SFR} \simeq 36$\,M$_\odot$\,yr$^{-1}$, based on near-infrared
photometry and the star formation model by \citet{bru93}. Hence the
SFR implied by our \cii\ upper limit is loosely consistent with that
inferred from the optical. However, given the approximations and
assumptions made here, it is hard to draw any firm conclusions.

\citet{luh03} find a deficit of \cii\ ($\lcii < 10^{-3}\,\lfir$) in a
sample of ULIRGs, and claim a negative correlation between \ratio\ and
$L_\nu(60\,\micron) / L_\nu(100\,\micron)$ for galaxies of all types (see
Fig.~5 in \citealt{luh03}). In order to compare our limit with
sources in this plot, we calculate
\[
L^\prime_\mathrm{FIR} = 1.26 \times 10^{12} \times 
\left[ 2.58\,L_\nu(60\,\micron)  +  L_\nu(100\,\micron) \right] ,
\]
with $L_\nu$ evaluated at rest-frame wavelengths (assuming a modified
blackbody) in W\,Hz$^{-1}$ and $L^\prime_\mathrm{FIR}$ in W (see
\citealt{hel88}). We find $L_{[C\,II]} / L^\prime_\mathrm{FIR} < 4.4
\times 10^{-3}$ and $L_\nu(60\,\micron) / L_\nu(100\,\micron) = 0.88$,
which is consistent with the published relation for normal
star-forming galaxies.  Nevertheless, firmer upper limits at similar
levels for a sample of high redshift star-forming galaxies will help
to determine if there is a difference in the role of \cii\ in the
early Universe.

\citet{bol04} recently reported an upper limit to \ratio\ in a
$z=6.42$ quasar, observed with the JCMT. They find $\ratio < 5 \times
10^{-4}$. While this limit is lower than that published here, we note
that a quasar is not a normal galaxy, and this source is $\sim 200$
brighter than ours, with $L_\mathrm{bol} \sim
10^{14}\,\mathrm{L}_\odot$. Our measurement, because of lensing
amplification, has allowed us to place a useful limit in a single
high-redshift star-forming galaxy. Clearly, this is a difficult
programme, and more sensitive measurements of both the \cii\ line
emission and the FIR are required to significantly improve the
constraint on \ratio.

\section*{Acknowledgments}

We thank the staff of the CSO for their help during observations and
analysis, and the referees for their insightful comments.  This work
was financially supported by the Natural Sciences and Engineering
Research Council of Canada.  The CSO is funded by the United States
National Science Foundation under contract AST 96-15025.  The SCUBA
data were obtained from the Canadian Astronomy Data Centre, which is
operated by the National Research Council of Canada's Herzberg
Institute of Astrophysics.

\bsp

\label{lastpage}

\end{document}